\documentclass{article}

\usepackage{microtype}
\usepackage{graphicx}
\usepackage{subcaption}
\usepackage{booktabs}
\usepackage[ruled, vlined]{algorithm2e}
\usepackage{amsmath, amssymb}
\usepackage{xcolor}
\usepackage{colortbl}
\usepackage{multirow}
\usepackage{makecell}
\usepackage{soul}
\usepackage{pdfpages}
\usepackage{afterpage}

\definecolor{commentgreen}{RGB}{0, 128, 0}

\SetCommentSty{mycommfont}

\usepackage{hyperref}

\usepackage[accepted]{preprint}
\usepackage[T1,OT1]{fontenc}
\hypersetup{pdfsubject={}, pdftitle={Mitigating SM-Level Load Imbalance in 3D Gaussian Splatting on GPUs}, pdfauthor={Junha Ryu, Hoi-Jun Yoo}}

\preprinttitlerunning{Mitigating SM-Level Load Imbalance in 3D Gaussian Splatting on GPUs}

\begin{document}

\twocolumn[
\preprinttitle{Mitigating SM-Level Load Imbalance in\texorpdfstring{\\}{ }3D Gaussian Splatting on GPUs}

\preprintsetsymbol{equal}{*}

\begin{preprintauthorlist}
\preprintauthor{Junha Ryu}{to}
\preprintauthor{Hoi-Jun Yoo}{to}
\end{preprintauthorlist}

\preprintaffiliation{to}{School of Electrical Engineering, Korea Advanced Institute of 
Science and Technology, Daejeon, South Korea}

\preprintemails{\{junha.ryu, hjyoo\}@kaist.ac.kr}
\preprintcorrespondingauthor{Hoi-Jun Yoo}{hjyoo@kaist.ac.kr}

\preprintkeywords{3D Gaussian Splatting, GPU, Load Balancing, Persistent Kernel}

\vskip 0.3in

\begin{abstract}
3D Gaussian Splatting (3DGS) enables photorealistic, real-time rendering on commodity GPUs and is increasingly adopted in generative world models and autonomous driving simulators. However, existing optimizations rely on algorithmic changes that often require retraining, or on specialized accelerators that lack flexibility across evolving 3DGS variants. In this paper, we identify a fundamental inefficiency in GPU-based 3DGS rendering: severe SM-level load imbalance caused by uneven workloads across screen tiles. This imbalance creates a long execution tail, leaving most SMs idle while a few process computationally intensive tiles, particularly in large-scale urban scenes. To address this, we present a software-defined GPU scheduling approach that mitigates the execution tail. Our approach uses a persistent kernel to schedule heavy tiles first and selectively chunks outlier tiles across multiple CTAs. A lightweight per-frame policy determines when and how much to chunk using tile statistics already available in the rendering pipeline. Our approach requires neither retraining nor specialized hardware and can be applied orthogonally to existing optimizations across diverse 3DGS variants. Averaged per scene, it accelerates alpha blending by up to $2.93\times$ on a large-scale urban scene, and by up to $5.6\times$ on its most imbalanced frames, while retaining the original alpha blending formulation.
\end{abstract}
]

\printAffiliationsAndNotice{}

\section{Introduction}
\label{sec:introduction}

3D Gaussian Splatting (3DGS)~\cite{3dgs} has emerged as a widely adopted 3D scene representation, enabling photorealistic rendering at real-time frame rates. Its ability to represent both reconstructed and generated 3D scenes has driven its adoption across diverse applications, including generative world models~\cite{lyra2,marble} and large-scale autonomous driving simulations~\cite{omnire}. For instance, NVIDIA Omniverse NuRec~\cite{nurec} reconstructs real-world driving scenes as 3DGS digital twins, which AlpaSim~\cite{alpasim} uses to evaluate vision-language-action (VLA) driving models such as Alpamayo~\cite{alpamayo}.

As these applications grow in scale and complexity, efficient 3DGS rendering becomes critical. However, current optimization efforts concentrate at the two ends of the stack: algorithm-level optimizations and specialized hardware acceleration. On the algorithmic front, research has primarily focused on reducing the overall Gaussian count through pruning and budget-constrained densification~\cite{mini_splatting,speedysplat,taming_3dgs}. However, these approaches are typically integrated into the training process, making them difficult to apply to already reconstructed 3D scenes without costly retraining. Moreover, these reductions do not inherently resolve the underlying architectural inefficiencies on general-purpose GPUs. On the hardware front, specialized accelerators~\cite{gscore,gsproc_isscc} offer high energy efficiency but lack the flexibility to support evolving 3DGS variants, such as 2DGS~\cite{2dgs} for better surface modeling and 3DGUT~\cite{3dgut} for diverse camera-model support. Furthermore, modern simulation environments such as AlpaSim necessitate the seamless integration of 3DGS with heterogeneous DNN workloads (e.g., VLA models). In these scenarios, the GPU remains the only viable platform as it provides both programmable cores for flexible rendering and tensor cores for high-throughput neural inference.

Despite the central role of GPUs, system-level software optimizations for maximizing GPU utilization remain relatively underexplored. Existing tile-based rasterizers assign each screen tile to a GPU Cooperative Thread Array (CTA), even though workloads vary substantially across tiles. Consequently, a few computationally intensive tiles can continue executing long after most others have finished, leaving many Streaming Multiprocessors (SMs) idle. This long execution tail is particularly pronounced in large-scale urban and driving scenes, hindering real-time rendering performance.

In this paper, we present a software-defined GPU scheduling approach that mitigates the execution tail in 3DGS rasterization. Our approach employs a persistent kernel to dispatch tiles in longest-processing-time-first (LPT) order based on their Gaussian counts, allowing heavy tiles to start early. Since scheduling alone cannot shorten exceptionally heavy tiles, it chunks these outliers into smaller units processed in parallel across multiple CTAs. By exploiting the composability of alpha blending, it preserves the original rendering formulation with negligible numerical differences, while saturation signaling restores early termination. To limit chunking overhead, it selectively chunks heavy tiles and co-executes them with LPT-scheduled tiles on separate CUDA streams. Finally, a lightweight online cost model determines how many tiles to chunk and how deeply for each frame, using Gaussian counts already available in the rendering pipeline.

Specifically, our contributions are as follows:
\begin{itemize}
\item \textbf{Quantitative Analysis of GPU Load Imbalance:} We characterize tile-level workload distributions and SM activity across diverse 3DGS variants and scenes, ranging from object-level benchmarks to city-scale environments. We demonstrate that a small number of computationally intensive tiles can dominate execution time, leaving most SMs idle and exposing a fundamental limitation of conventional GPU rasterization.
\item \textbf{LPT Scheduling via Persistent Kernel:} We introduce workload-aware tile scheduling through a persistent kernel. By dispatching tiles in descending order of their projected Gaussian counts, our scheduler prioritizes heavy tiles and reduces the execution tail caused by late CTA dispatch.
\item \textbf{Depth-wise Chunking with Saturation Signaling:} LPT scheduling changes when a tile executes, not how long it takes, leaving exceptionally heavy tiles on the critical path. We exploit the composability of alpha blending to partition these tiles into depth-contiguous chunks that execute in parallel across multiple CTAs. To avoid redundant computation, we introduce saturation signaling that restores early termination across chunks while retaining the original alpha blending formulation.
\begin{figure}[t]
    \centering
    \includegraphics[width=0.97\linewidth, trim={6mm 9.5mm 7.8mm 6.5mm}, clip]{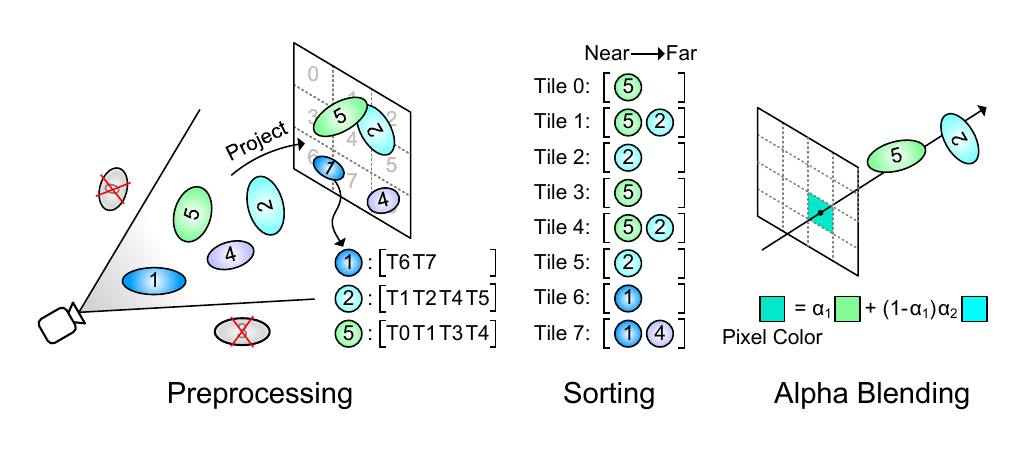}
    \caption{Common rendering pipeline of 3DGS and its variants.}
    \label{fig:3dgs}
\end{figure}
\item \textbf{Selective Chunking and Stream Co-execution:} Chunking every tile incurs unnecessary overhead, particularly on relatively balanced workloads. We therefore chunk only the heaviest tiles and co-execute their chunks with LPT-scheduled tiles on separate CUDA streams. This allows the GPU hardware scheduler to dynamically allocate SM resources between the two workloads.
\item \textbf{Online Per-Frame Chunking Policy:} We develop a lightweight cost model that balances execution-tail reduction against chunking overhead. Using per-tile Gaussian counts, the policy selects both the number of tiles to chunk and the chunk depth for each frame, including the option to disable chunking. This enables online adaptation without exhaustive configuration searches or scene-specific tuning.
\item \textbf{Performance Validation:} We evaluate our approach across diverse 3DGS workloads, from object-level benchmarks to large-scale urban and driving scenes, achieving up to $2.93\times$ speedup in alpha blending execution. Our approach supports multiple 3DGS variants, including 2DGS, 3DGUT, Hier-3DGS~\cite{hier_3dgs}, and CityGaussianV2~\cite{citygaussianv2}. It can also be applied orthogonally to existing GPU rasterization optimizations, including SwiftGS~\cite{swiftgs} and HiGS~\cite{higs}, without modifying scene representations or requiring retraining.
\end{itemize}

\section{Background}
\label{sec:background}

\begin{figure*}[t]
    \centering
    \begin{minipage}[c]{0.48\linewidth}
        \centering
        \includegraphics[width=\linewidth]{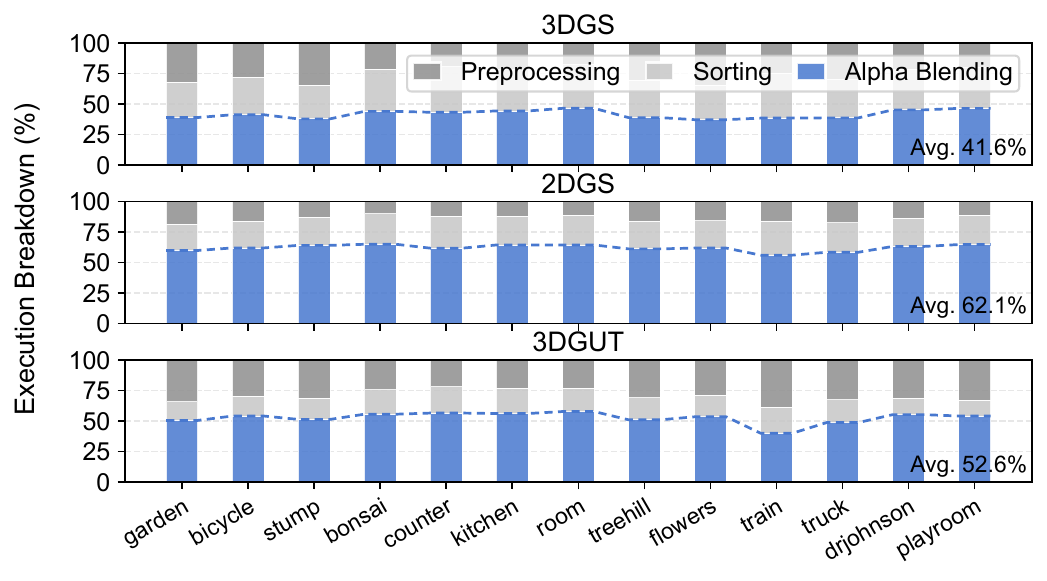}
    \end{minipage}\hfill
    \begin{minipage}[c]{0.48\linewidth}
        \centering
        \includegraphics[width=0.97\linewidth]{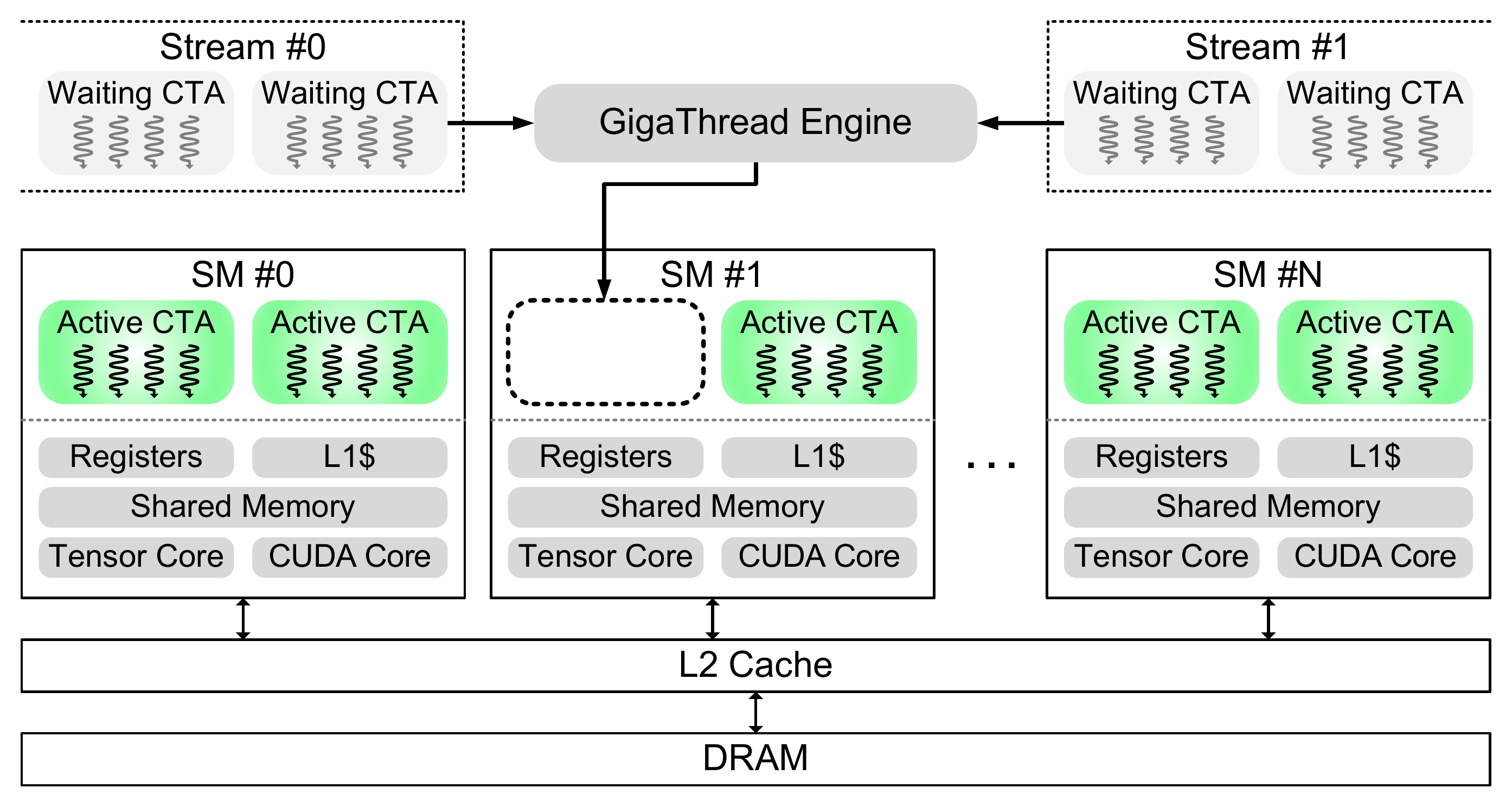}
    \end{minipage}\\[2pt]
    \begin{minipage}[t]{0.48\linewidth}
        \caption{Rendering-time breakdown across 3DGS, 2DGS, and 3DGUT.}
        \label{fig:time_breakdown}
    \end{minipage}\hfill
    \begin{minipage}[t]{0.48\linewidth}
        \caption{GPU execution model and CTA scheduling across CUDA streams and SMs.}
        \label{fig:gpu_model}
    \end{minipage}
    \vspace{-6pt}
\end{figure*}

\subsection{3D Gaussian Splatting and Its Variants}

3DGS represents a 3D scene as a collection of anisotropic 3D Gaussians, each defined by a center $\mu \in \mathbb{R}^3$, a covariance $\Sigma$ parameterized by scale and rotation, an opacity $o \in [0,1]$, and a color. As illustrated in Figure~\ref{fig:3dgs}, the rendering pipeline consists of three stages: preprocessing, sorting, and alpha blending. During preprocessing, Gaussians outside the camera's field of view are culled, and the remaining Gaussians are projected onto the image plane. Given a viewing transformation $W$ and projection Jacobian $J$, the projected 2D covariance $\Sigma'$ is:
\begin{equation}\label{eq:proj}\Sigma' = JW\Sigma W^TJ^T.\end{equation}

The screen is partitioned into tiles, and each projected Gaussian with center $\mu' \in \mathbb{R}^2$ is assigned to all overlapping tiles. During sorting, Gaussians within each tile are ordered front to back by camera-space depth. During alpha blending, each tile accumulates Gaussian contributions to its pixels. For a pixel coordinate $x \in \mathbb{R}^2$, the alpha value of Gaussian $i$ is:
\begin{equation}
    \alpha_i = o_i \cdot \exp\left(-\frac{1}{2}(x-\mu')^T 
    \Sigma'^{-1} (x-\mu')\right).
\end{equation}
Then, the final pixel color $C$ is accumulated as:
\begin{equation}
\label{eq:blend}
    C = \sum_{i=1}^{N} c_i \alpha_i T_i, \qquad T_i = \prod_{j=1}^{i-1}(1-\alpha_j).
\end{equation}
Here, $c_i$ denotes the color of Gaussian $i$, and $T_i$ is the remaining transmittance before processing it. Since transmittance decreases monotonically, a pixel stops blending once $T_i$ falls below $10^{-4}$, where subsequent contributions become negligible. This early termination allows a tile to finish before processing all its assigned Gaussians.

Beyond conventional 3DGS, 2DGS~\cite{2dgs} uses oriented planar disks and computes $\alpha_i$ via ray--splat intersection for accurate surface reconstruction. 3DGUT~\cite{3dgut} replaces the linearized projection of Eq.~\eqref{eq:proj} with the unscented transform to support distorted cameras and secondary rays. Although these variants differ in how $\alpha_i$ is computed, they retain the same tile-based alpha blending of Eq.~\eqref{eq:blend}. As shown in Figure~\ref{fig:time_breakdown}, alpha blending accounts for a substantial fraction of rendering time across these variants, making it a common target for optimization.

\subsection{GPU Architecture and CTA Scheduling}
\begin{figure*}[t]
    \centering
    \includegraphics[width=\linewidth]{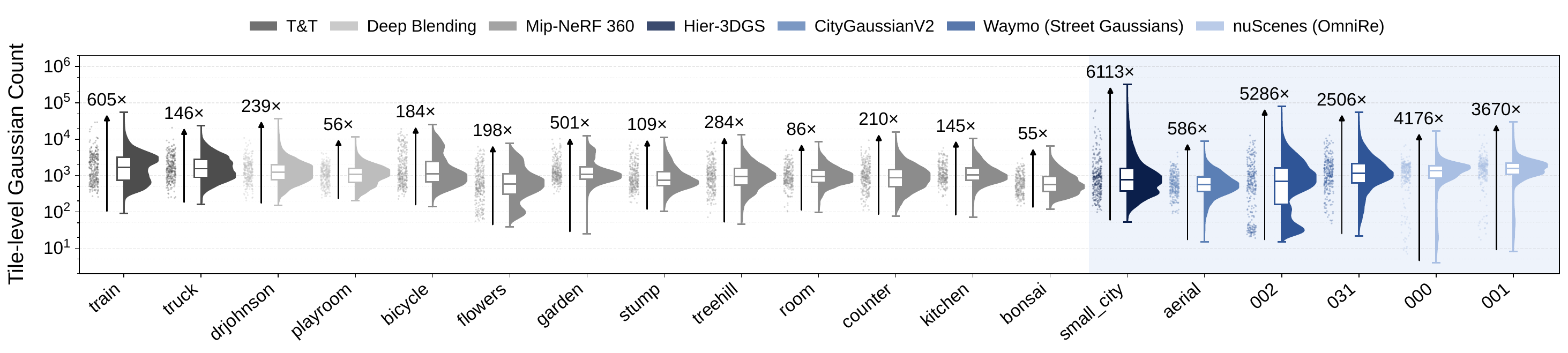}\par
    \vspace{-8pt}
    \caption{Distribution of per-tile Gaussian counts across small-scale and large-scale scenes, the latter shaded blue. Each raincloud plot shows the density (violin), interquartile range (box), and per-tile samples (dots). Driving scenes use the front camera.}
    \label{fig:gaussian_count}
    \vspace{-8pt}
\end{figure*}

Figure~\ref{fig:gpu_model} illustrates the GPU execution model relevant to 3DGS rasterization. Modern GPUs consist of multiple SMs, each equipped with CUDA cores, registers, and shared memory. When a kernel is launched, its threads are organized into CTAs, each consisting of one or more 32-thread warps. The hardware CTA scheduler (GigaThread Engine in NVIDIA GPUs) assigns CTAs to available SMs based on resource requirements. Each CTA executes on a single SM, while multiple CTAs can reside on the same SM concurrently. Kernels launched in separate CUDA streams may also execute concurrently when GPU resources permit.

In conventional tile-based 3DGS rasterization, each $16 \times 16$-pixel tile is mapped to one 256-thread CTA, with each thread computing the final color of one pixel. Since the hardware scheduler operates at CTA granularity, it cannot redistribute a tile's workload across multiple SMs once its CTA is scheduled. Consequently, large variations in tile workloads can leave a few long-running CTAs executing while other SMs become idle. This rigid mapping between screen tiles and CTAs limits GPU utilization, motivating more flexible workload scheduling and parallelization.
\section{Challenges and Opportunities}
\label{sec:motivation}

\subsection{Workload Imbalance in Tile-Based Rendering}
\label{sec:tile_imbalance}

Prior work accelerates 3DGS rendering by reducing Gaussian processing through pruning or tighter bounding volumes \cite{speedysplat, flashgs}. While these optimizations reduce overall computation, they do not directly address how the remaining workload is distributed across GPU CTAs. To characterize this distribution, we measure per-tile Gaussian counts across 19 scenes spanning standard benchmarks (Tanks \& Temples, Deep Blending, and Mip-NeRF 360) and large-scale urban and driving scenes. Figure~\ref{fig:gaussian_count} reveals highly skewed workload distributions. Standard benchmarks exhibit maximum-to-minimum count ratios of $55$--$605\times$, while large-scale scenes reach over $6{,}100\times$. However, these counts are only a proxy for blending work, since early termination can skip Gaussians. More importantly, a large skew alone does not imply low GPU utilization.

\subsection{SM Utilization Across 3DGS Variants}
\label{sec:sm_util}
\begin{figure}[t]
    \centering
    \begin{minipage}[b]{\linewidth}
        \centering
        \includegraphics[width=\linewidth, trim={0mm 7mm 7mm 5mm}, clip]{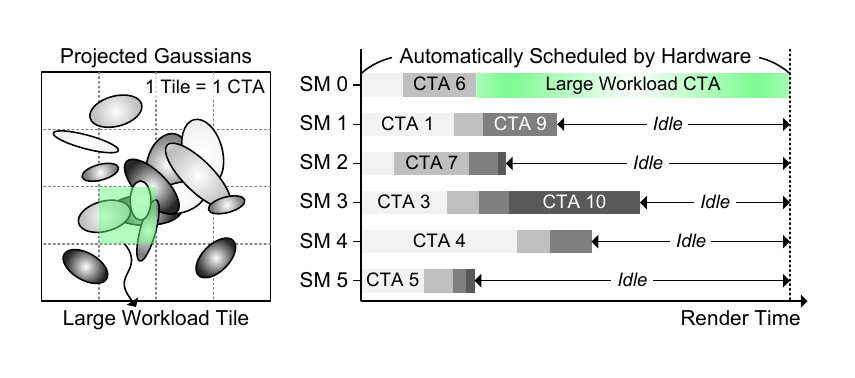}
        \centerline{(a)}
        \label{fig:sm_imbalance}
    \end{minipage}
    \\[-1.5ex]
    \begin{minipage}[b]{\linewidth}
        \centering
        \includegraphics[width=\linewidth]{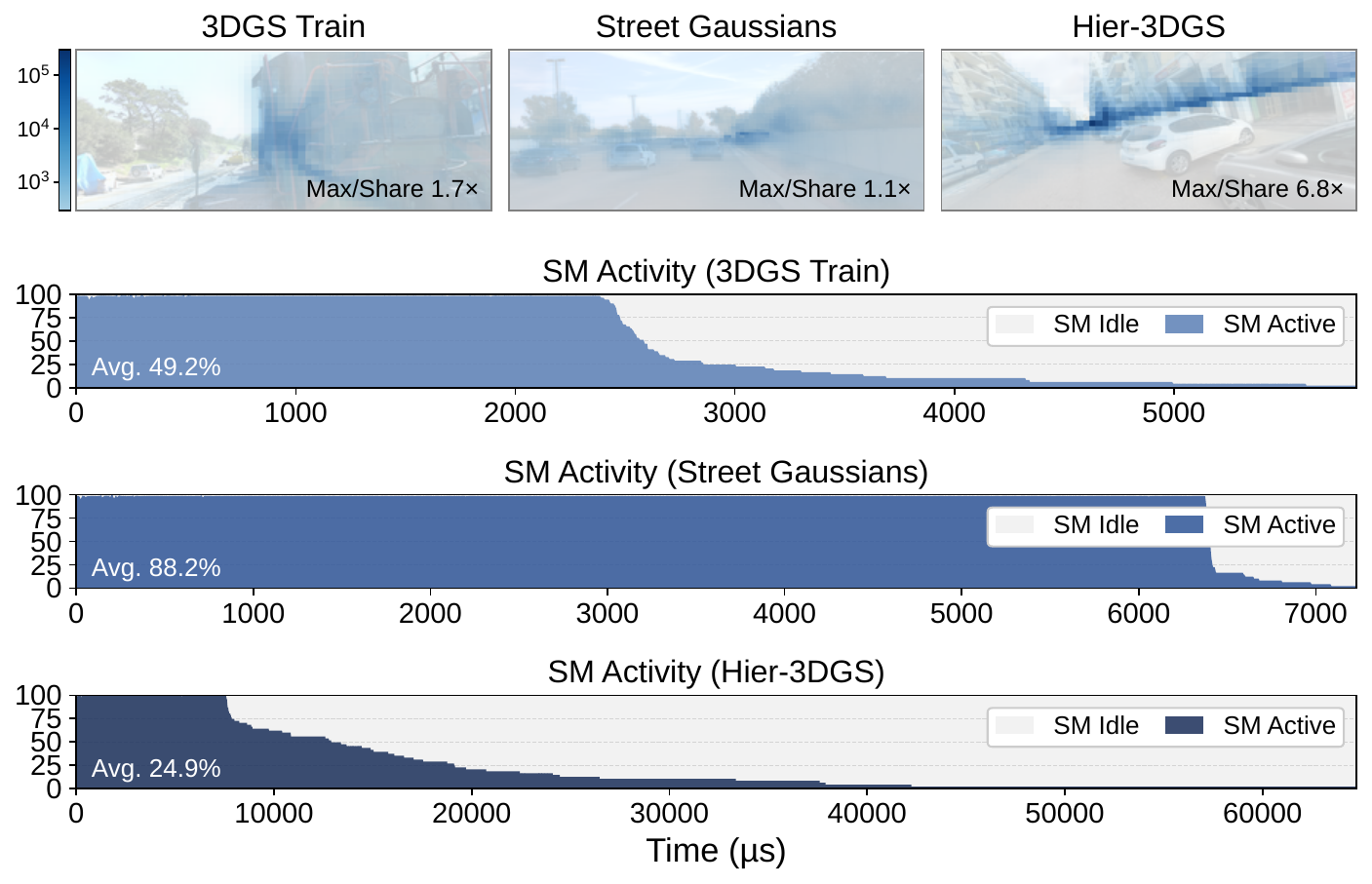}
        \centerline{(b)}
        \label{fig:pm_sampling}
    \end{minipage}
    \par\vspace{-3ex}
    \caption{(a) Execution tail caused by tile imbalance. (b) Per-tile Gaussian counts and corresponding SM activity timelines for representative frames, sampled with Nsight Compute~\cite{nsight_compute}.}
    \label{fig:sm_imbalance_sampling}
    \vspace{-12pt}
\end{figure}
We now examine when this skew translates into GPU underutilization. Even with highly uneven tiles, the GPU remains busy as long as enough CTAs are available; a long execution tail arises only when the heaviest tiles are large relative to the GPU's parallel capacity and keep executing after most other CTAs have completed. To quantify this effect, let $N_t$ denote the projected Gaussian count of tile $t$, $N_{\mathit{total}} = \sum_t N_t$, and $K$ the number of CTAs the GPU can execute concurrently. Under an ideal distribution, each CTA would process a share of $N_{\mathit{total}}/K$ Gaussians. Since conventional rasterization assigns each tile to a single CTA, a tile much heavier than this share can dominate the critical path, as illustrated in Figure~\ref{fig:sm_imbalance_sampling}(a). We use the ratio of the heaviest tile to this share as a proxy for the execution tail.

Figure~\ref{fig:sm_imbalance_sampling}(b) compares this ratio with measured SM activity. In Street Gaussians~\cite{street_gaussians}, the heaviest tile is only $1.1\times$ the per-CTA share, and average SM activity remains high at $88.2\%$. In contrast, the 3DGS \textit{train} scene exhibits a ratio of $1.7\times$, with average SM activity of $49.2\%$. The imbalance is even more pronounced in Hier-3DGS, where the heaviest tile reaches $6.8\times$ the share and average SM activity drops to $24.9\%$. In the latter two, only a few SMs remain active near the end of the frame.

\begin{figure}[t]
    \centering
    \includegraphics[width=\linewidth]{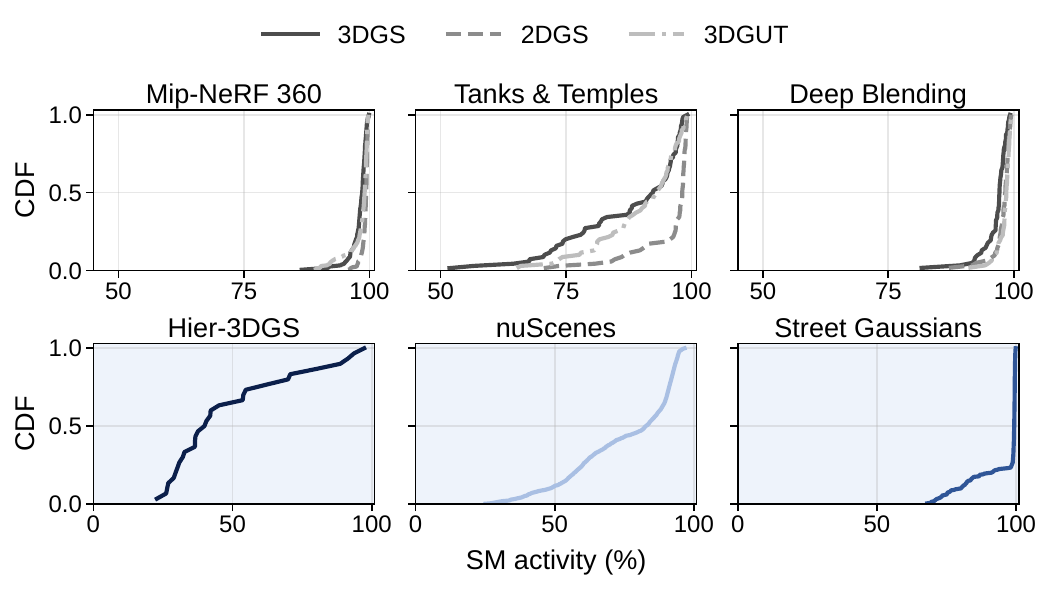}
    \vspace{-26pt}
    \caption{CDF of per-frame SM activity across 3DGS variants and datasets (front camera for driving scenes), measured with Nsight Compute.}
    \label{fig:sm_utilization_cdf}
    \vspace{-10pt}
\end{figure}

Figure~\ref{fig:sm_utilization_cdf} extends this analysis to six dataset categories. Here, SM activity is the fraction of elapsed cycles during which an SM has at least one resident warp; it reflects SM idleness rather than instruction throughput or occupancy. Standard benchmarks mostly sustain high utilization, with some underutilized frames. Among large-scale scenes, Hier-3DGS exhibits low utilization across many frames, while the nuScenes front camera falls below 80\% in nearly half of its frames. In contrast, Street Gaussians stays highly utilized in most frames despite substantial tile skew. Underutilization thus varies across scenes and even across frames of the same scene, so load balancing must account for both the heaviest tiles and the available GPU parallelism rather than applying one strategy to every frame.

\subsection{Opportunities}
\label{sec:opportunities}

\paragraph{CTA Scheduling via Persistent Kernel.}
The hardware CTA scheduler is unaware of tile workloads, so a heavy tile may start late and finish after all others. However, the per-tile Gaussian count available after sorting can serve as a workload estimate. A persistent kernel~\cite{ptb} allows a fixed pool of CTAs to repeatedly fetch tiles from a software-managed work queue, enabling workload-aware execution ordering. Persistent-kernel approaches have also been explored in LLM workloads, including FlashAttention-3~\cite{flashattention3} and FlashInfer~\cite{flashinfer}, while Mirage Persistent Kernel (MPK)~\cite{mpk} extends persistent execution to an entire inference pass.

\paragraph{Alpha Blending Composability.}
Scheduling changes when a tile executes, but cannot shorten the workload of a tile assigned to a single CTA. Alpha blending's composability removes this limitation. Consider partitioning a tile's depth-ordered Gaussian list before Gaussian $m$. The front segment (Gaussians $1, \ldots, m-1$) produces a partial color $C_A$ and remaining transmittance $T_m$ as defined in Eq.~\eqref{eq:blend}, while the back segment independently computes $C_B$ assuming unit initial transmittance. The results can be combined as
\begin{equation}
\label{eq:compose}
    C = C_A + T_m \cdot C_B.
\end{equation}
This identity allows a tile's depth-ordered workload to be divided into contiguous chunks that execute independently across multiple CTAs, while retaining the original alpha blending formulation. Since 2DGS, 3DGUT, and large-scale variants such as Hier-3DGS and CityGaussianV2~\cite{citygaussianv2} share this compositing structure, the same opportunity extends across diverse 3DGS renderers.

\section{Method}
\label{sec:method}

Our approach mitigates GPU execution tails through workload-aware tile scheduling and selective intra-tile parallelism. We first use a persistent kernel to dispatch tiles in longest-processing-time-first (LPT) order based on their Gaussian counts (\S\ref{sec:method-sched}). For outlier tiles that remain bottlenecks under LPT, we introduce depth-wise chunking with saturation signaling to parallelize their execution across multiple CTAs while reducing redundant computation (\S\ref{sec:method-split}). We co-execute chunked and LPT-scheduled tiles on separate CUDA streams (\S\ref{sec:method-policy}), while a lightweight per-frame cost model determines when and how much to chunk (\S\ref{sec:method-policy-auto}). These optimizations preserve the original alpha blending formulation without modifying scene representations or requiring specialized hardware.

\subsection{LPT Scheduling via Persistent Kernel}
\label{sec:method-sched}

\begin{figure}[!t]
    \centering
    \begin{minipage}[b]{\linewidth}
        \centering
        \includegraphics[width=\linewidth, trim={0mm 7mm 7mm 5mm}, clip]{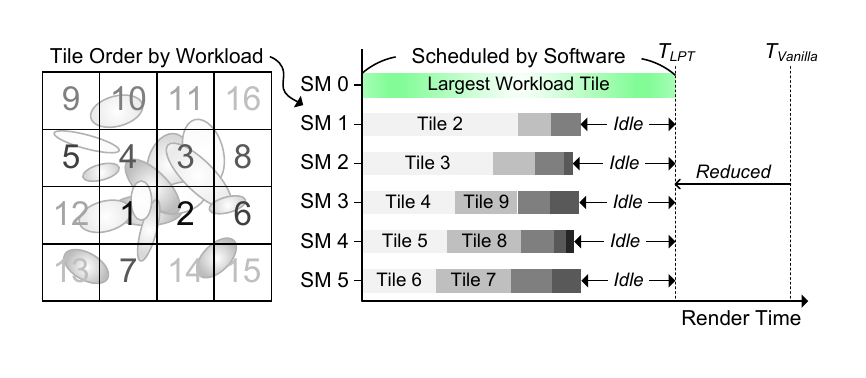}
        \centerline{(a)}
        \label{fig:ptb_sorting}
    \end{minipage}
    \\[-1.5ex]
    \begin{minipage}[b]{\linewidth}
        \centering
        \includegraphics[width=\linewidth]{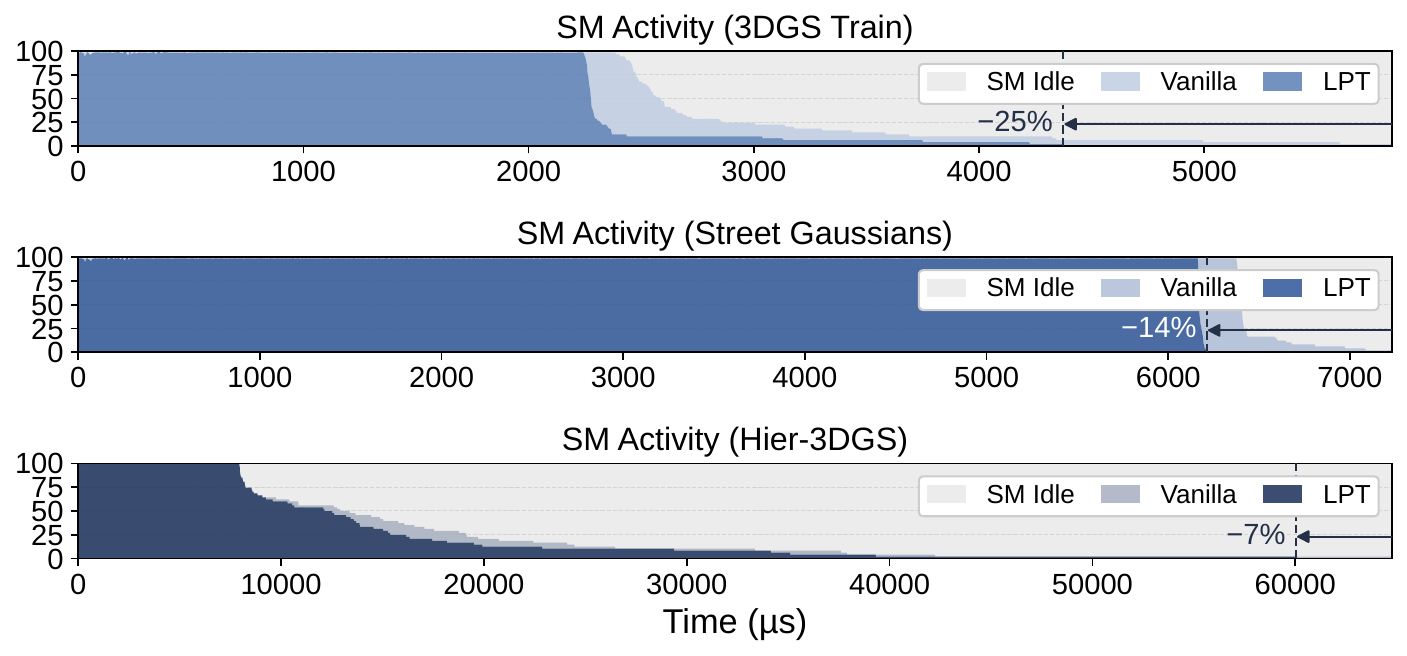}
        \centerline{(b)}
        \label{fig:pm_sampling_ptb}
    \end{minipage}
    \\[-1.5ex]
    \begin{minipage}[b]{\linewidth}
        \centering
        \includegraphics[width=\linewidth]{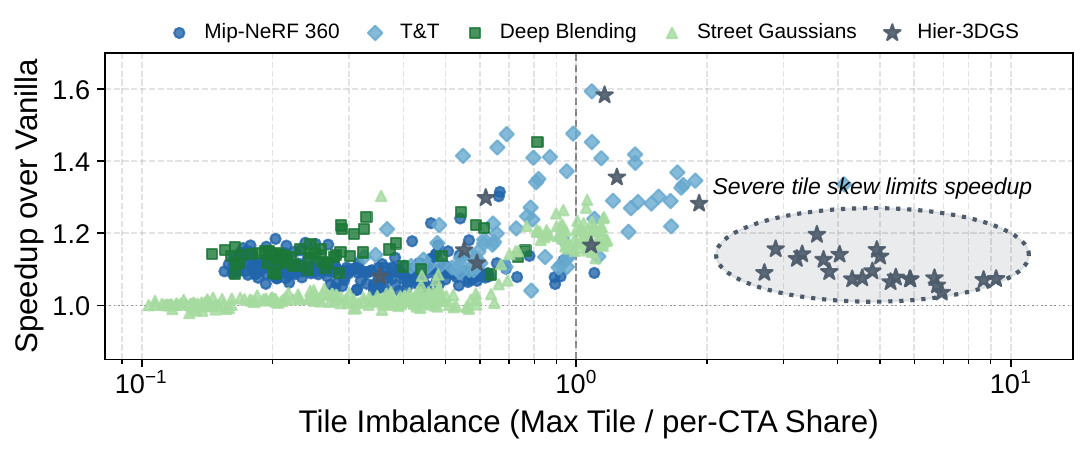}
        \centerline{(c)}
        \label{fig:ptb_lpt_speedup}
    \end{minipage}
    \par\vspace{-3ex}
    \caption{(a) LPT scheduling via a persistent kernel. (b) SM activity before and after LPT scheduling for the frames in Figure~\ref{fig:sm_imbalance_sampling}(b). (c) LPT speedup over Vanilla versus the heaviest tile relative to the per-CTA share.}
    \label{fig:ptb_lpt}
\end{figure}

Conventional 3DGS rasterizers launch one CTA per tile and leave the dispatch order to the hardware scheduler, which is unaware of tile workloads. As a result, a heavy tile may start late and become the execution tail. To control the execution order, we instead launch a persistent kernel with $K$ CTAs, the number the GPU can run concurrently. Each CTA then repeatedly takes the next tile from a software-managed work queue $Q$ until the queue is empty, with one thread per CTA advancing the queue through a single \textit{atomicAdd}.

\paragraph{LPT ordering.}
We order $Q$ so that the heaviest tiles are dispatched first. To estimate the workload of tile $t$, we use its projected Gaussian count $N_t$, which is directly available from the tile's range in the sorted Gaussian list. We then radix-sort the tile indices in descending order of $N_t$ without moving the Gaussian data. This counting and sorting takes less than $1\%$ of the alpha blending kernel time on average across the three workloads in Figure~\ref{fig:ptb_lpt}(b). As illustrated in Figure~\ref{fig:ptb_lpt}(a), the heavy tile thus starts first while the other CTAs process lighter tiles, so the kernel finishes at $T_{\mathit{LPT}}$ instead of $T_{\mathit{Vanilla}}$.

\paragraph{Effectiveness and limitation.}
Figure~\ref{fig:ptb_lpt}(b) shows that LPT shortens the alpha blending kernel of the profiled frame by $25\%$ on 3DGS \textit{train} and $14\%$ on Street Gaussians, but by only $7\%$ on Hier-3DGS. This is because the heaviest tile of Hier-3DGS is $6.8\times$ the per-CTA share and still runs long after the other tiles finish, even when dispatched first. Figure~\ref{fig:ptb_lpt}(c) shows the same trend across frames. LPT achieves $1.1$--$1.6\times$ speedup on many standard benchmark frames. In contrast, Street Gaussians frames, whose heaviest tile mostly stays below the per-CTA share, have little tail to remove and gain only $1.06\times$ on average. At the other extreme, most Hier-3DGS frames, whose heaviest tile far exceeds the share, gain only $1.10\times$. In other words, LPT changes when a tile executes but not how long it takes, so shortening such outlier tiles requires parallelism within a tile, which we introduce next.

\subsection{Depth-wise Chunking with Saturation Signaling}
\label{sec:method-split}

\begin{figure}[t]
    \centering
    \begin{minipage}[b]{\linewidth}
        \centering
        \includegraphics[width=\linewidth, trim={5mm 5mm 5mm 7mm}, clip]{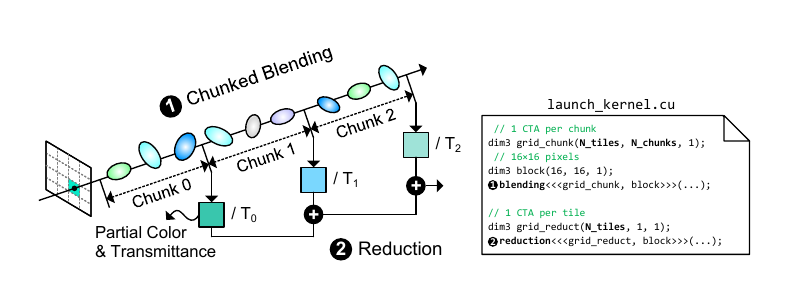}
        \\[1.3ex]
        \hspace*{3mm}\includegraphics[width=0.93\linewidth, trim={20.5bp 16bp 16.5bp 18.3bp}, clip]{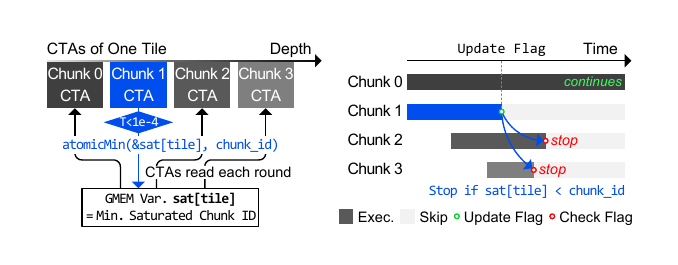}
        \centerline{(a)}
    \end{minipage}
    \\[-0.4ex]
    \begin{minipage}[b]{\linewidth}
        \centering
        \includegraphics[width=\linewidth, trim={0mm 0mm 0mm 1.5mm}, clip]{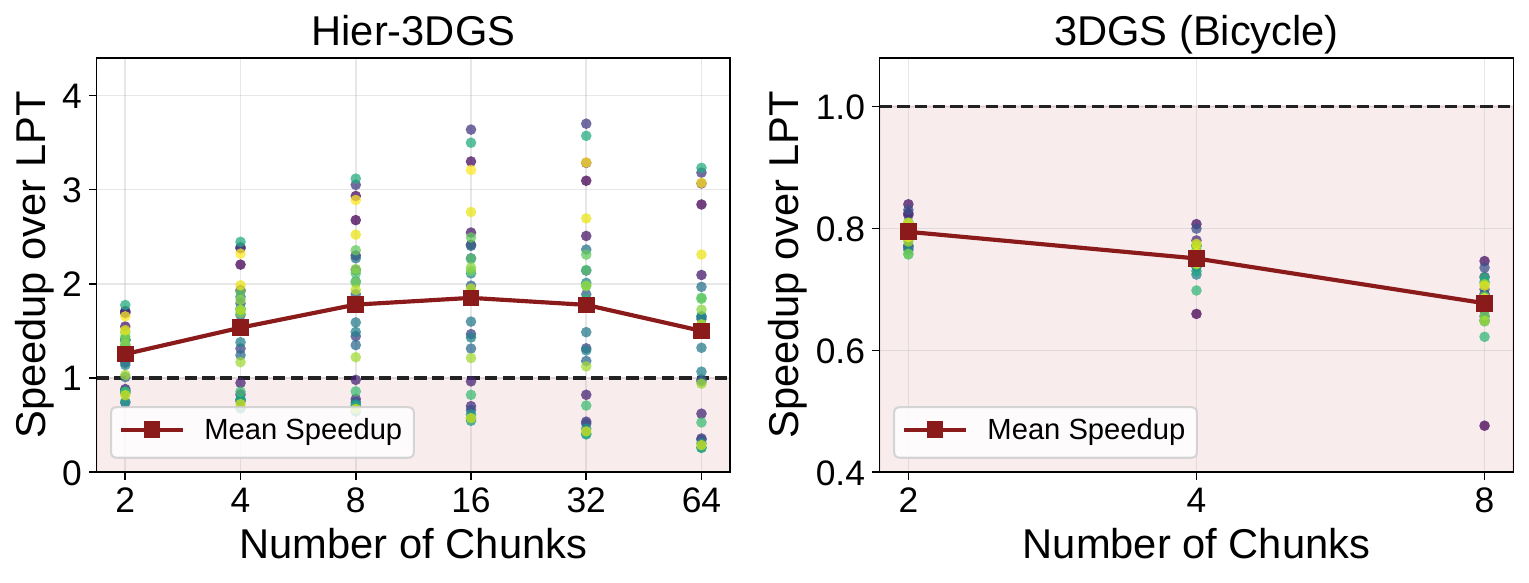}
        \centerline{(b)}
    \end{minipage}
    \par\vspace{-1.5ex}
    \caption{(a) Depth-wise chunking and saturation signaling. (b) Per-frame speedup of Chunking-only over LPT-only versus the number of chunks.}
    \label{fig:split_overview}
\end{figure}

To shorten an exceptionally heavy tile, we partition its depth-ordered Gaussian list into $k$ contiguous chunks, each processed by a separate CTA. Each chunk independently performs alpha blending with unit initial transmittance and produces a partial color $C_p$ and transmittance $T_p$ for each pixel. A reduction kernel then combines these partial results in depth order:
\begin{equation}
\label{eq:merge}
    C = \sum_{p=1}^{k} C_p \prod_{q=1}^{p-1} T_q.
\end{equation}
This extends Eq.~\eqref{eq:compose} to $k$ chunks, allowing their blending computations to execute independently across multiple SMs (Figure~\ref{fig:split_overview}(a)). Similar to Flash-Decoding~\cite{dao2023flashdecoding}, this approach parallelizes a long sequential computation across CTAs and combines partial results, although 3DGS requires per-pixel color and transmittance for composition.

\paragraph{Saturation signaling.}
Independent chunk execution disrupts early termination because each chunk starts with unit transmittance, unaware of the opacity accumulated by preceding chunks. Consequently, later chunks may process Gaussians that the original sequential renderer would skip. We address this through saturation signaling, which communicates early-termination information across independently executing CTAs (Figure~\ref{fig:split_overview}(a)). Each chunked tile maintains a global-memory flag recording the earliest chunk whose pixels have all reached the transmittance threshold. Upon saturation, a chunk updates the flag using \textit{atomicMin}. Later chunks check it once per 256-Gaussian batch and stop when an earlier chunk has saturated. This is safe because transmittance only decreases in depth order. If an earlier chunk alone has driven every pixel below the threshold, the true transmittance before any later chunk is even lower, so the sequential renderer would also have stopped before the skipped Gaussians. On 3DGS \textit{train}, saturation signaling accelerates Chunking-only execution by $1.21\times$ on average and up to $1.66\times$ per frame.

\paragraph{Benefits and overheads.}
Depth-wise chunking shortens outlier tiles but incurs additional global-memory traffic for partial results and a reduction pass. Figure~\ref{fig:split_overview}(b) evaluates this trade-off by chunking every tile and comparing against LPT-only execution. On Hier-3DGS, Chunking-only achieves an average $1.9\times$ speedup over LPT at $k=16$, as distributing outlier workloads across CTAs reduces the execution tail. Conversely, on the relatively balanced 3DGS \textit{bicycle} scene, it achieves only $0.68$--$0.79\times$ the performance of LPT on average across chunk counts. Since \textit{bicycle} has no extremely heavy tile, chunking barely shortens the frame, while every chunked tile still pays the extra memory traffic and reduction. Chunking therefore helps only for the heavy tiles that bound the frame, which motivates chunking them selectively rather than every tile.

\subsection{Selective Chunking and Co-execution}
\label{sec:method-policy}

\begin{figure}[!t]
    \centering
    \begin{minipage}[b]{\linewidth}
        \centering
        \includegraphics[width=\linewidth, trim={0mm 7mm 6mm 5mm}, clip]{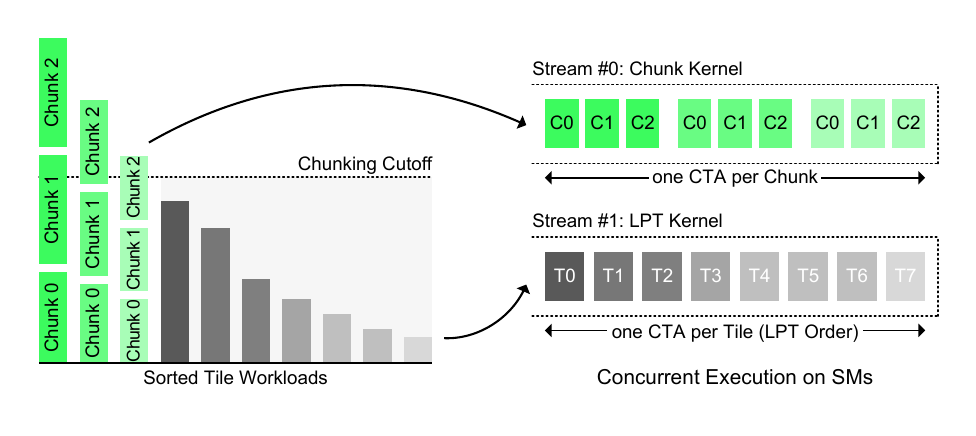}
        \centerline{(a)}
    \end{minipage}
    \\[-0.4ex]
    \begin{minipage}[b]{\linewidth}
        \centering
        \includegraphics[width=\linewidth]{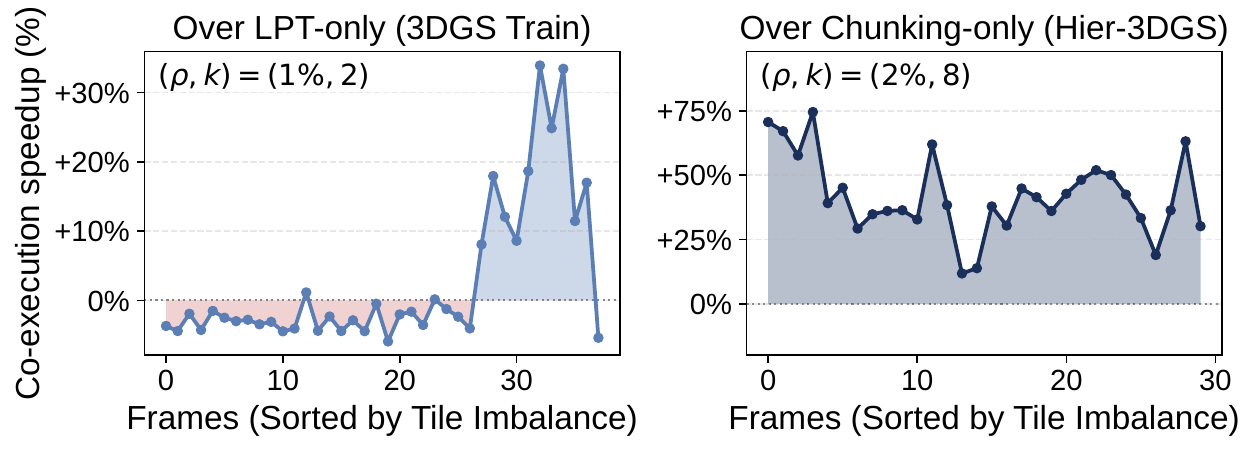}
        \centerline{(b)}
    \end{minipage}
    \par\vspace{-1.5ex}
    \caption{(a) Stream co-execution of chunk and LPT kernels. (b) Per-frame speedup of co-execution over LPT-only (left) and Chunking-only (right).}
    \label{fig:coexec-frames}
\end{figure}

LPT avoids late dispatch of heavy tiles, whereas chunking shortens individual outliers. Since neither performs best across all workload distributions, we combine both within a frame, chunking the heaviest tiles while scheduling the remainder in LPT order.

\paragraph{Stream co-execution.}
Given a heavy-tile ratio $\rho$ and chunk count $k$, we select the heaviest fraction $\rho$ of tiles from the existing LPT order and partitions each into $k$ chunks. The remaining tiles are assigned to the persistent LPT kernel without additional sorting. As shown in Figure~\ref{fig:coexec-frames}(a), the chunk kernel assigns one CTA per chunk, while the persistent kernel dispatches the remaining tiles in LPT order. The kernels execute on separate CUDA streams, allowing their CTAs to share SM resources as hardware scheduling and resource availability permit. After both kernels complete, a reduction kernel combines the partial results, with dependencies enforced through CUDA events.

\paragraph{Effectiveness of co-execution.}
Figure~\ref{fig:coexec-frames}(b) compares co-execution with LPT-only and Chunking-only using the best fixed $(\rho, k)$ configuration for each workload. On Hier-3DGS \textit{small\_city}, co-execution outperforms Chunking-only by $12$--$75\%$ across frames. On highly imbalanced 3DGS \textit{train} frames, it is up to $34\%$ faster than LPT-only. However, chunking even the heaviest $1\%$ of tiles slows relatively balanced 3DGS \textit{train} frames by up to $6\%$. The best fixed configuration also differs between 3DGS \textit{train} $(1\%, 2)$ and Hier-3DGS $(2\%, 8)$. These results motivate adapting both the number of chunked tiles and the chunk count to each frame.

\subsection{Per-Frame Chunking Policy}
\label{sec:method-policy-auto}

Since tile imbalance varies from frame to frame, the best $(h, k)$ must be chosen at runtime without searching over configurations. Chunking too few tiles leaves outliers on the critical path, whereas chunking too many adds unnecessary overhead. We therefore use a lightweight cost model to select two parameters on the fly for each frame: the number of heaviest tiles to chunk, $h$, and the number of chunks per tile, $k$. The policy uses only the Gaussian counts already available from LPT sorting.

\paragraph{Cost estimation model.}
Let $N_t$ denote the projected Gaussian count of tile $t$, indexed so that $N_1 \geq N_2 \geq \cdots$, and let $N_{\mathit{total}} = \sum_t N_t$. The model balances two competing terms. First, chunking $h$ tiles into $k$ chunks each adds an overhead of $\lambda hk$, where $\lambda$ is a constant modeling the per-chunk cost of storing and reducing partial results. Distributed across $K$ concurrent CTA workers, this yields an average-workload term of $(N_{\mathit{total}} + \lambda hk)/K$. Second, execution time is constrained by the longest remaining work unit. After chunking the $h$ heaviest tiles, we estimate its size as
\[
    \ell(h,k) = \max\left(\frac{N_1}{k},\, N_{h+1}\right),
\]
where the two terms represent the largest chunked workload and the heaviest unchunked tile. Adding a penalty $\beta\ell$ for residual scheduling imbalance gives
\begin{equation}
\label{eq:makespan}
    \hat{T}(h,k) = \max\left(\frac{N_{\mathit{total}} + \lambda hk}{K},\, \ell\right) + \beta\ell.
\end{equation}
$\hat{T}$ is a heuristic cost expressed in Gaussian-equivalent work units, rather than a direct prediction of measured latency.

\begin{figure}[t]
    \centering
    \includegraphics[width=\linewidth]{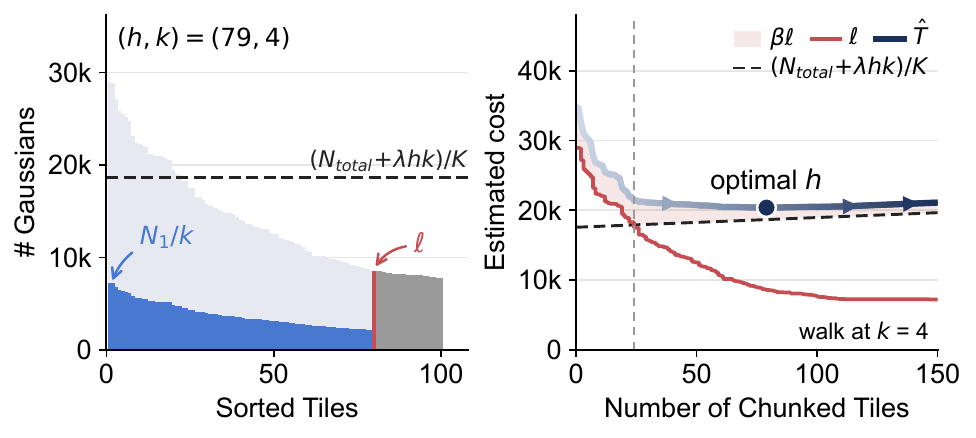}\par
    \vspace{-8pt}
    \caption{Per-frame chunking cost estimation on a 3DGS \textit{train} frame. Left: Workload distribution after chunking. Right: Estimated cost versus the number of chunked tiles.}
    \label{fig:cost-model}
\end{figure}

\paragraph{Choosing $h$ and $k$.}
For each candidate $k$, the policy evaluates increasing $h$ in descending workload order. Each additional chunked tile can reduce $\ell$ but raises the average-workload term by $\lambda k/K$. Figure~\ref{fig:cost-model} illustrates this trade-off for $k=4$ on a 3DGS \textit{train} frame. Chunking the first $24$ tiles reduces the critical-path term; beyond this point, further chunking lowers the $\beta\ell$ penalty until the estimated cost reaches its minimum at $h=79$. The policy searches $k \in \{2, 4, 8, 16\}$ and selects the $(h^*, k^*)$ minimizing $\hat{T}$, including the no-chunking option ($h=0$). To account for workload overestimation caused by early termination, chunking is enabled only when the estimated cost is at least $\epsilon = 10\%$ lower than LPT-only.

\begin{figure}[t]
    \centering
    \includegraphics[width=\linewidth]{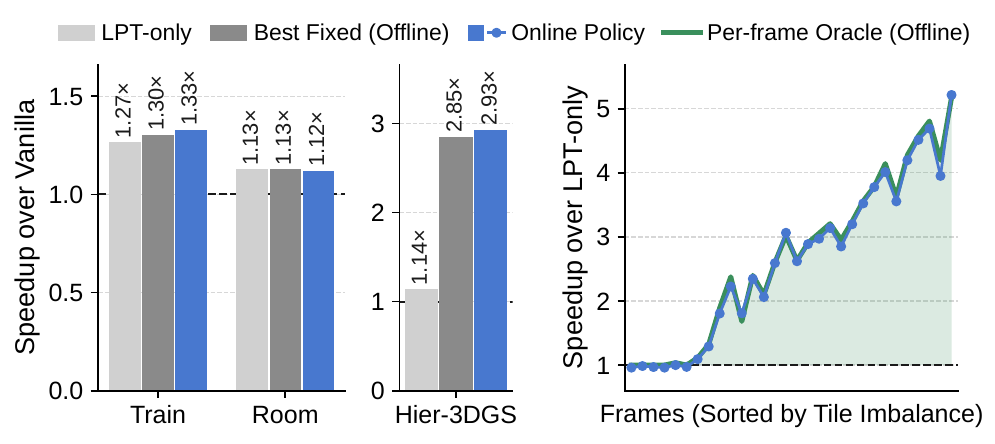}\par
    \vspace{-8pt}
    \caption{Effectiveness of the per-frame chunking policy.}
    \label{fig:policy-effect}
\end{figure}

\paragraph{Effectiveness of the online policy.}
Figure~\ref{fig:policy-effect} compares our policy with two offline baselines that are chosen in hindsight after measuring a set of candidate $(\rho, k)$ configurations on every frame. The best fixed baseline applies the single configuration with the highest average speedup to all frames of a scene, and the per-frame oracle picks the fastest configuration for each frame individually. Neither is available at runtime. Even so, our policy outperforms the best fixed baseline by $2.0\%$ on 3DGS \textit{train} and $2.7\%$ on Hier-3DGS. On 3DGS \textit{room}, it avoids chunking and remains within $0.6\%$ of LPT-only. On Hier-3DGS, it performs within $2.0\%$ of the per-frame oracle on average without any offline configuration search.

\section{Evaluation}
\label{sec:eval}

\subsection{Experimental Setup}
\label{sec:eval-setup}

We evaluate our approach on an NVIDIA GB10 (48 SMs) and an NVIDIA RTX A6000 (84 SMs).
We render all thirteen scenes of Mip-NeRF 360, Tanks \& Temples and Deep
Blending with 3DGS, four of them with 2DGS and 3DGUT, \textit{train} and
\textit{truck} with SwiftGS~\cite{swiftgs} and HiGS~\cite{higs},
\textit{small\_city} with Hier-3DGS, and the MatrixCity \textit{aerial} and
\textit{street} scenes with the 2DGS-based CityGaussianV2~\cite{citygaussianv2}.
Through gsplat~\cite{gsplat}, we render the front camera of $15$ driving
sequences from nuScenes, Argoverse~2, PandaSet and Waymo reconstructed by
OmniRe~\cite{omnire}, eight camera views of NVIDIA NuRec~\cite{nurec} scenes
with 3DGUT, and three scenes generated by Lyra 2.0. We compare \textit{Vanilla}, the
original rasterizer; \textit{LPT-only} (\S\ref{sec:method-sched});
\textit{Chunking-only}, which chunks every tile (\S\ref{sec:method-split}); and
\textit{Ours}, which combines LPT dispatch with selective chunking under the
per-frame policy (\S\ref{sec:method-policy}). We report the alpha blending
speedup over Vanilla and image quality in PSNR, SSIM and LPIPS.

\subsection{Overall Performance}
\label{sec:eval-overall}

\begin{figure*}[t]
    \centering
    \includegraphics[width=\linewidth]{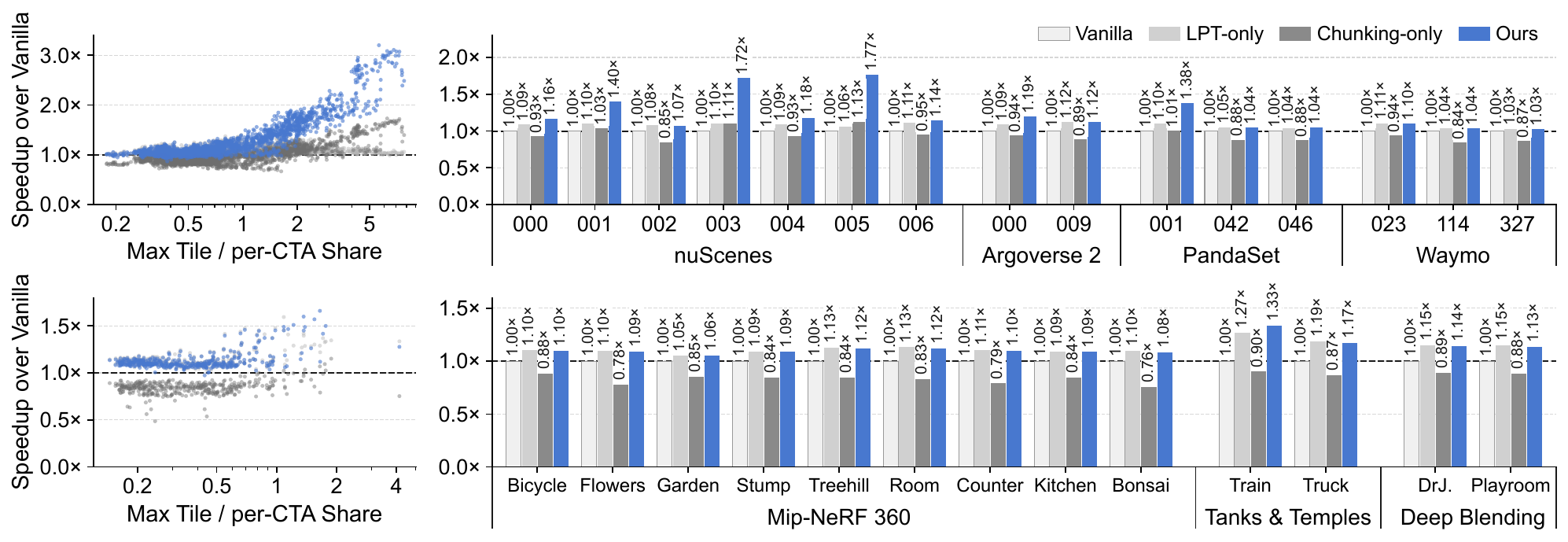}\par
    \vspace{-8pt}
    \caption{Speedup over Vanilla on driving scenes (top) and standard 3DGS benchmarks (bottom), per frame versus tile imbalance (left) and per scene (right).}
    \label{fig:overall}
\end{figure*}

\paragraph{Rendering Performance.}
As shown in Figure~\ref{fig:overall}, our approach achieves an average speedup of
$1.23\times$ on the driving scenes and $1.12\times$ on the standard
benchmarks, and up to $3.20\times$ on individual frames. On the standard
benchmarks, most of the gain comes from LPT scheduling. On the most imbalanced driving scenes, selective chunking is essential. For
example, on nuScenes scene 005, our approach reaches $1.77\times$, whereas LPT-only
achieves $1.06\times$. Chunking-only, in contrast,
adds overhead on relatively balanced scenes, which the per-frame policy
avoids.

\definecolor{qgrp}{HTML}{F3F5F9}
\begin{table}[t]
\centering
\caption{Quality difference of Chunking-only from Vanilla.}
\label{tab:quality}
\footnotesize
\setlength{\tabcolsep}{4.5pt}
\begin{tabular}{@{}llc rrr@{}}
\toprule
Renderer & Dataset & \#Scene & $\Delta$PSNR & $\Delta$SSIM & $\Delta$LPIPS \\
\midrule
\rowcolor{qgrp} & M360 & 9 & +0.0001 & 0.0000 & 0.0000 \\
\rowcolor{qgrp} & T\&T & 2 & $-$0.0001 & 0.0000 & 0.0000 \\
\rowcolor{qgrp}\multirow{-3}{*}{3DGS} & DB & 2 & $-$0.0004 & 0.0000 & 0.0000 \\
 & M360 & 3 & +0.0002 & 0.0000 & 0.0000 \\
\multirow{-2}{*}{2DGS} & T\&T & 1 & $-$0.0001 & 0.0000 & 0.0000 \\
\rowcolor{qgrp} & M360 & 3 & +0.0002 & 0.0000 & 0.0000 \\
\rowcolor{qgrp}\multirow{-2}{*}{3DGUT} & T\&T & 1 & +0.0001 & 0.0000 & 0.0000 \\
\multirow{-1}{*}{Hier-3DGS} & City & 1 & $-$0.0015 & 0.0000 & 0.0000 \\
\rowcolor{qgrp}\multirow{-1}{*}{Street GS} & Waymo & 2 & $-$0.0002 & 0.0000 & 0.0000 \\
\bottomrule
\end{tabular}
\end{table}

\paragraph{Rendering Quality.}
LPT scheduling changes only the dispatch order of tiles, whereas depth-wise
chunking reorders the floating-point accumulation within a tile. Table~\ref{tab:quality}
therefore compares Vanilla with Chunking-only, which chunks every tile and thus
changes the accumulation order the most. Across $24$ scenes, the differences are at most $0.0015$\,dB in
PSNR, $2.4\times10^{-5}$ in SSIM and $7\times10^{-5}$ in LPIPS. On every frame
of the driving and NuRec workloads, our output also stays above
$76$\,dB PSNR relative to the original renderer.

\subsection{Generality and Applicability}
\label{sec:eval-generality}

\begin{figure}[t]
    \centering
    \includegraphics[width=\linewidth]{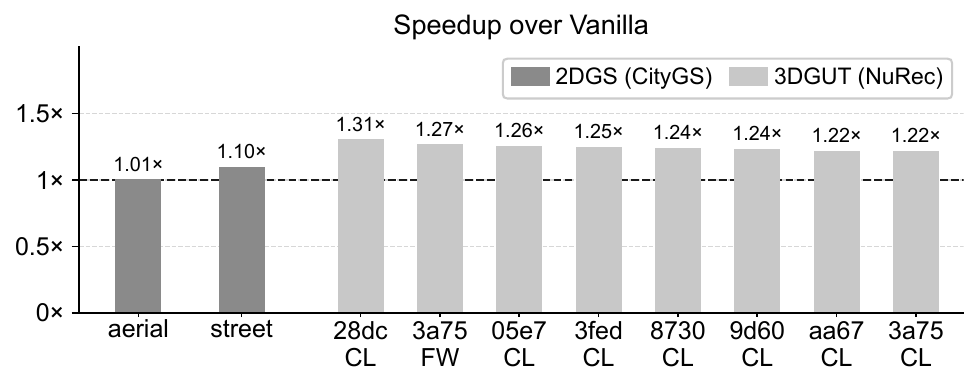}\par
    \vspace{-8pt}
    \caption{Speedup of our approach over Vanilla with CityGaussianV2 on MatrixCity and 3DGUT on NuRec views (CL: cross-left, FW: front-wide).}
    \label{fig:variants}
\end{figure}

\begin{figure}[t]
    \centering
    \includegraphics[width=\linewidth]{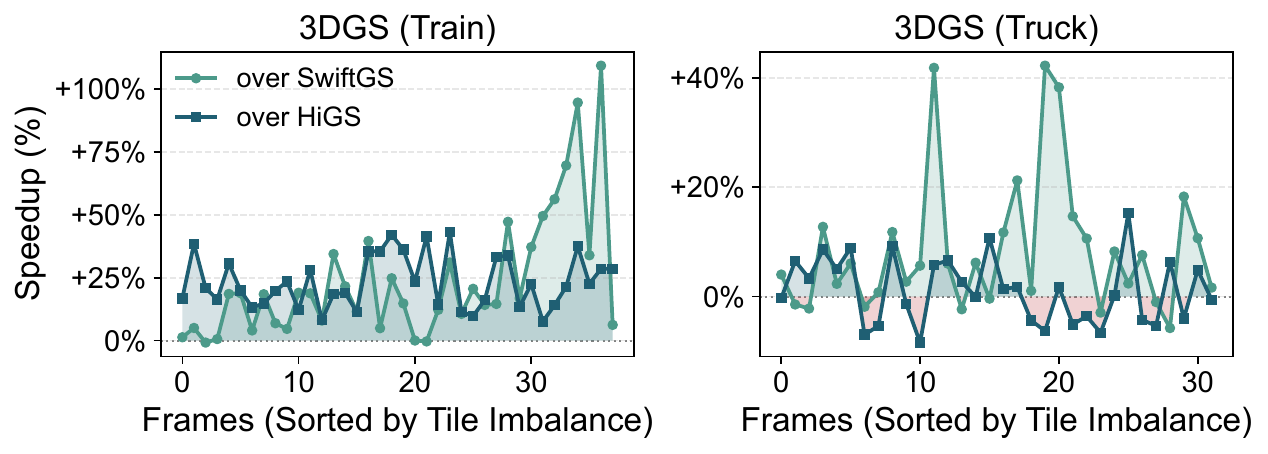}\par
    \vspace{-8pt}
    \caption{Per-frame speedup of our approach over SwiftGS and HiGS.}
    \label{fig:optraster}
\end{figure}

\paragraph{3DGS Variants.}
Our approach applies to any renderer built on tile-based alpha blending, without
modifying the scene representation or the trained model. Beyond 3DGS
(Figure~\ref{fig:overall}) and Hier-3DGS ($2.93\times$ on average and up to $5.64\times$ per frame,
Figure~\ref{fig:policy-effect}), it achieves $1.22$--$1.31\times$ on the
NuRec views rendered with 3DGUT and $1.10\times$ on \textit{street} with
the 2DGS-based CityGaussianV2, while staying on par with Vanilla ($1.01\times$) on
\textit{aerial} (Figure~\ref{fig:variants}).

\paragraph{Combining with Optimized Rasterizers.}
SwiftGS~\cite{swiftgs} reduces the per-Gaussian blending cost, and
HiGS~\cite{higs} renders fixed-size batches of each macro-tile's Gaussian list
on separate CTAs and merges them in a post-blending pass. Neither orders work by
cost, and HiGS blends every batch even after its macro-tile saturates. We apply
our approach unchanged to SwiftGS. For HiGS, whose lists are already chunked, we add
saturation signaling across batches, dispatch earlier batches first, and render
single-batch macro-tiles without the post-blending pass. As shown in
Figure~\ref{fig:optraster}, our approach achieves $1.24\times$ over SwiftGS (up to
$2.09\times$) and $1.23\times$ over HiGS ($1.08$--$1.43\times$ on every frame)
on \textit{train}, and $1.08\times$ and $1.01\times$ on the less skewed
\textit{truck}.

\begin{figure}[t]
    \centering
    \includegraphics[width=\linewidth]{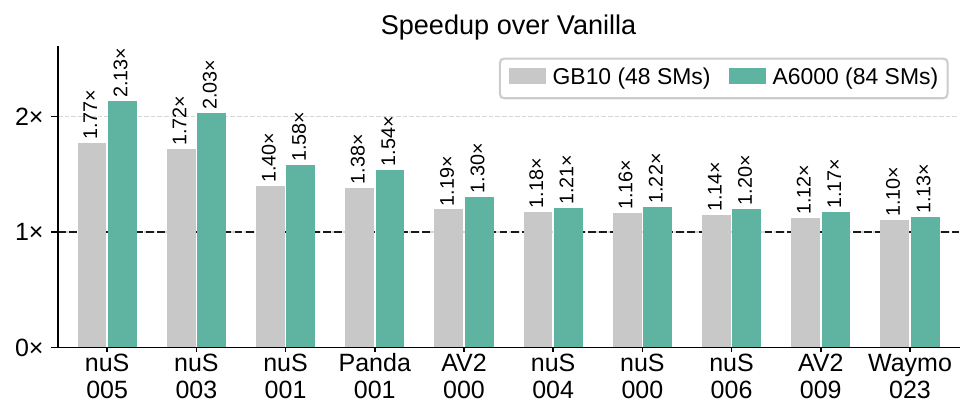}\par
    \vspace{-8pt}
    \caption{Speedup of our approach over Vanilla on the GB10 (48 SMs) and RTX A6000 (84 SMs).}
    \label{fig:gpu_arch}
\end{figure}

\paragraph{GPU Architectures.}
The impact of the execution tail depends on the number of SMs: a GPU with more
SMs finishes the light tiles sooner, while a heavy tile still takes just as
long. We therefore examine how effective our approach is on GPUs with different SM
counts, comparing an RTX A6000 with $84$ SMs and the GB10 with $48$ SMs. As shown in Figure~\ref{fig:gpu_arch}, on
ten driving scenes, it achieves $1.13$--$2.13\times$ ($1.45\times$ on
average) on the A6000, compared with $1.10$--$1.77\times$ ($1.32\times$) on the
GB10. Our approach therefore becomes more beneficial
as GPUs scale to more SMs.

\afterpage{\afterpage{%
\begin{figure}[t]
    \centering
    \includegraphics[width=\linewidth]{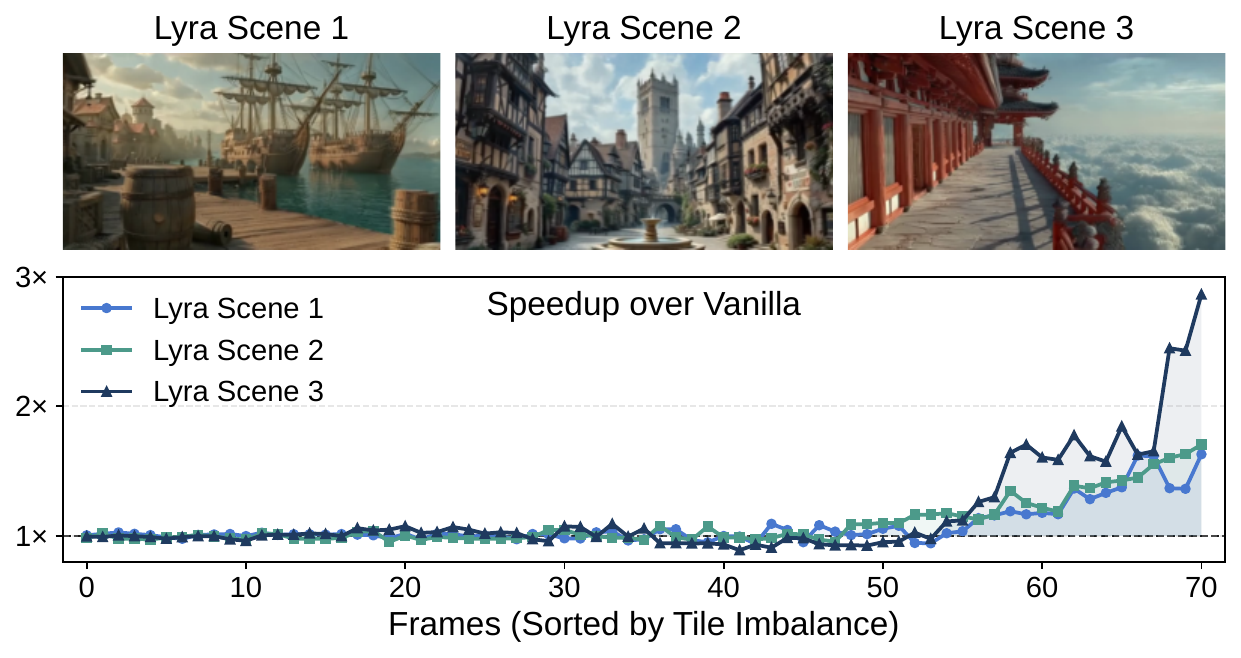}\par
    \vspace{-8pt}
    \caption{Per-frame speedup of our approach over Vanilla on Lyra 2.0-generated scenes.}
    \label{fig:lyra}
\end{figure}
\begin{figure}[t]
    \centering
    \includegraphics[width=\linewidth]{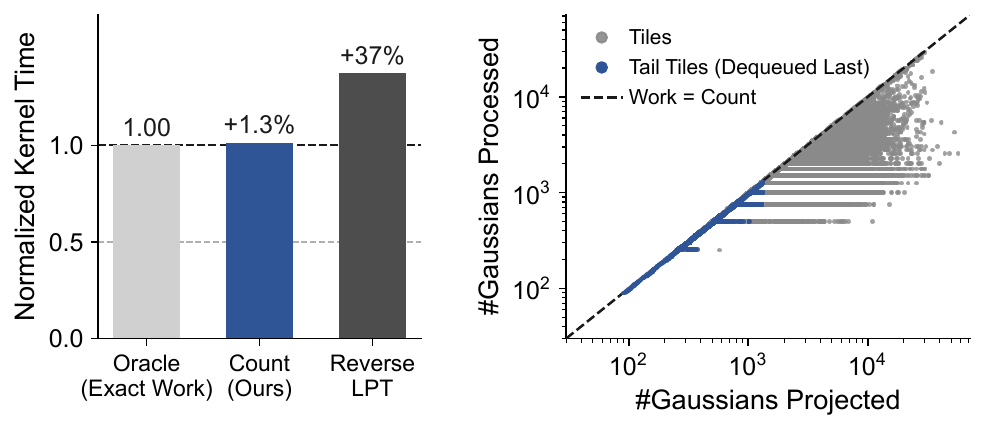}\par
    \vspace{-8pt}
    \caption{Gaussian count as an LPT workload proxy on 3DGS \textit{train}. Left: kernel time by tile order. Right: processed versus projected Gaussians per tile.}
    \label{fig:oracle_order}
\end{figure}
\begin{figure}[t]
    \centering
    \includegraphics[width=\linewidth]{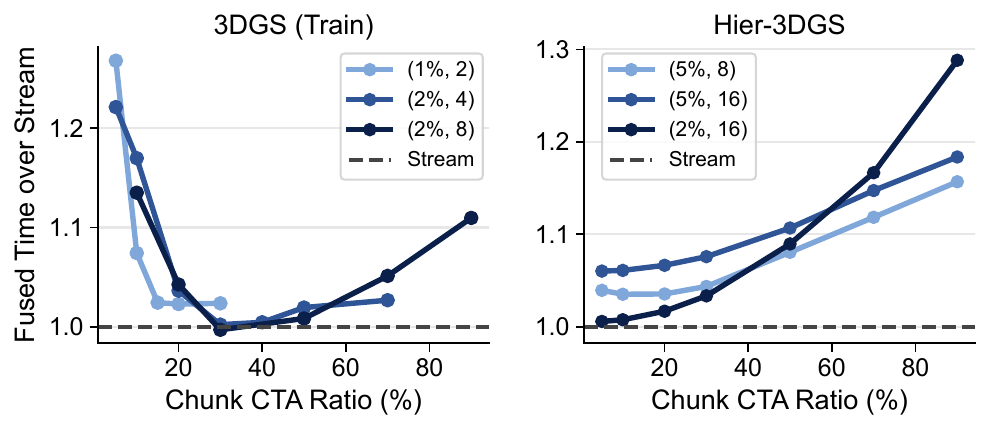}\par
    \vspace{-8pt}
    \caption{Fused-kernel latency normalized to stream co-execution versus chunk-CTA ratio.}
    \label{fig:fused_vs_stream}
\end{figure}
\begin{figure}[t]
    \centering
    \includegraphics[width=\linewidth]{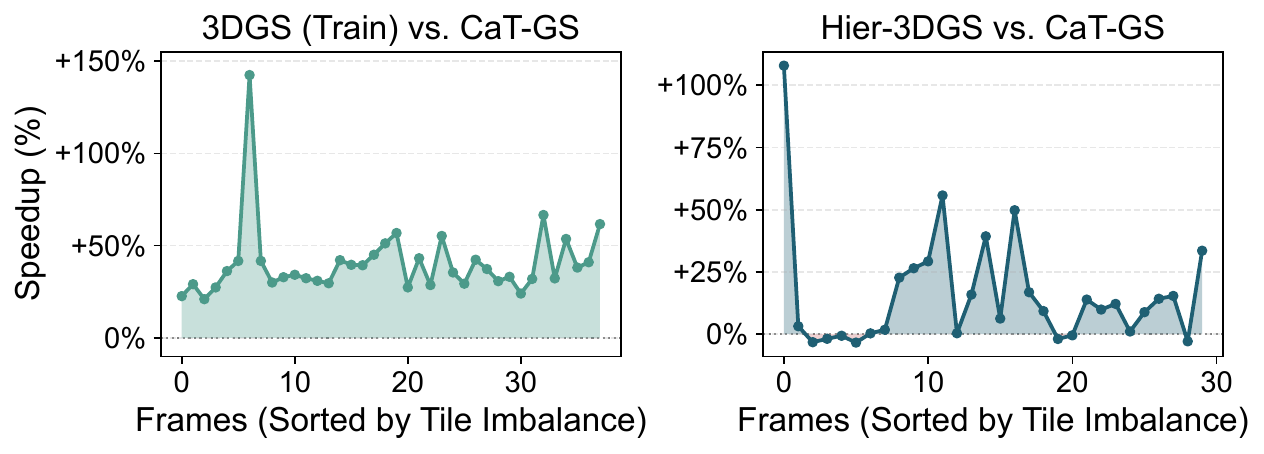}\par
    \vspace{-8pt}
    \caption{Per-frame speedup of our approach over CaT-GS.}
    \label{fig:catgs}
\end{figure}}}

\paragraph{World-Model-Generated Scenes.}
Generative world models such as Lyra 2.0~\cite{lyra2} synthesize 3D scenes directly
as Gaussians rather than reconstructing them from captured images, and their
Gaussian density varies widely across viewpoints. As shown in Figure~\ref{fig:lyra}, our approach achieves $1.07$--$1.17\times$
on average over the three scenes and up to $2.86\times$ on the most imbalanced
frames, whereas LPT-only achieves only $1.01$--$1.06\times$. The per-frame
speedup closely follows the tile imbalance, indicating that its benefit
depends on the workload distribution rather than on how the scene is
constructed.

\subsection{Design Alternatives}
\label{sec:eval-design}

\paragraph{Gaussian Count as a Workload Proxy.}
LPT orders tiles by their projected Gaussian count, which ignores early
termination. An offline oracle that orders tiles by the number of Gaussians
actually processed, measured from a prior execution of the same frame, is only
$1.3\%$ faster, whereas reversing the order increases execution time by $37\%$
(Figure~\ref{fig:oracle_order}). Since early termination only reduces work,
projected Gaussian counts provide conservative workload estimates, enabling
near-oracle LPT scheduling without runtime profiling.

\paragraph{Stream Co-execution versus a Fused Kernel.}
Alternatively, the chunk and LPT workloads can share one persistent kernel
whose CTAs are partitioned between them, with work stealing when either queue
empties. As shown in Figure~\ref{fig:fused_vs_stream}, its latency is
sensitive to the fraction of CTAs assigned to chunks, whose best value varies
across workloads. Even with workload-specific tuning, fusion reduces latency by less than $1\%$, whereas a suboptimal partition increases it by up to $29\%$.
Stream co-execution thus achieves comparable performance without
partition tuning by leaving SM allocation to the hardware scheduler.

\paragraph{Comparison with Mean-Threshold Tile Splitting.}
CaT-GS~\cite{catgs} splits every tile exceeding twice the mean Gaussian count
into mean-sized chunks and dispatches them after the regular tiles. This
threshold depends only on the workload distribution, not on how many CTAs the
GPU executes concurrently, so it can split far more tiles than needed. Each chunk adds
partial-result traffic and reduction, and without saturation signaling, chunks
keep blending Gaussians that sequential rendering would skip. Our approach instead
chunks only when its cost model predicts a shorter critical path. As a result, it is $40.5\%$ and $16.0\%$ faster on average
on 3DGS \textit{train} and Hier-3DGS, respectively, with a maximum slowdown of
$3.4\%$ on any frame (Figure~\ref{fig:catgs}).

\section{Limitations and Conclusion}
\label{sec:conclusion}

\paragraph{Limitations.}
Our approach targets the alpha blending stage of inference-time rendering, and its
benefit scales with the tile imbalance of each frame. On balanced workloads,
such as the Waymo and two of the PandaSet sequences, the execution tail is short
and the room for improvement is limited ($1.03$--$1.10\times$ in
Figure~\ref{fig:overall}); there, the per-frame policy rarely chunks, and
our approach performs close to LPT-only without slowing down Vanilla. The policy uses a single set of cost-model
constants and projected Gaussian counts across all evaluated workloads.
Calibrating these constants per GPU or learning them online could
further narrow the gap to the per-frame oracle, which we leave to future work.

\paragraph{Conclusion.}
This paper identified SM-level load imbalance across screen tiles as a
fundamental source of inefficiency in GPU-based 3DGS rendering. We presented
a software-defined GPU scheduling approach that dispatches heavy
tiles first through a persistent kernel, chunks outlier tiles across multiple
CTAs with saturation signaling, and co-executes the chunks with LPT-scheduled
tiles under a lightweight per-frame policy. Across 3DGS variants, optimized
rasterizers, and GPU architectures, our approach accelerates alpha blending by up
to $2.93\times$ on a large-scale urban scene and by up to $5.6\times$ on its most
imbalanced frames, without retraining, modifying scene
representations, or changing the alpha blending formulation. As 3DGS
rendering scales to city-scale driving simulation and generative world
models, our approach provides a practical path to efficient rendering on
general-purpose GPUs.

\bibliography{main}
\bibliographystyle{preprint}

\end{document}